\documentclass[reprint,amsmath,amssymb,aps]{revtex4-1}    
\usepackage{graphicx}
\usepackage{xcolor}
\graphicspath{{./figures/}}

\begin{document}

\title{Regulating dynamics through intermittent interactions}%
\author{Shiva Dixit }
\thanks{These authors contributed equally to this work}
\email[\\ Corresponding author email: ]{manaojaravind@iitb.ac.in}
\author{Manaoj Aravind}
\thanks{These authors contributed equally to this work}
\email[\\ Corresponding author email: ]{manaojaravind@iitb.ac.in}
\author{P. Parmananda}
\affiliation{Department of Physics, Indian Institute of Technology Bombay, Powai, Mumbai $400$ $076$, India.}

\begin{abstract}
    In this letter, we experimentally demonstrate an efficient scheme to regulate the behaviour of coupled nonlinear oscillators through dynamic control of their interaction. It is observed that introducing intermittency in the interaction term as a function of time or the system state, predictably alters the dynamics of the constituent oscillators. Choosing the nature of the interaction - attractive or repulsive, allows for either suppression of oscillations or stimulation of activity. Two parameters $\Delta$ and $\tau$, that reign the extent of interaction among subsystems are introduced.  They serve as a harness to access the entire range of possible behaviours from fixed points to chaos. For fixed values of system parameters and coupling strength, changing $\Delta$ and $\tau$  offers fine control over the dynamics of coupled subsystems. We show this experimentally using coupled Chua's circuits and elucidate their behaviour for a range of coupling parameters through detailed numerical simulations. 
\end{abstract}

\date{\today}%
\maketitle

\section{Introduction}\label{Introduction}

Dynamics ranging from fixed points to chaos exist in nature~\cite{lakshmanan1996chaos} and mechanisms to regulate such dynamics are crucial for practical applications. Coupled nonlinear oscillators serve as a prototype to model many real-world systems~\cite{winfree1980geometry}; examples include the climate~\cite{scheffer2020critical}, population dynamics in ecosystems~\cite{scheffer2001catastrophic} and financial markets~\cite{scheffer2001catastrophic}. Coupling among dynamical systems gives rise to fascinating emergent behaviour like synchronization~\cite{pikovsky2001synchronisation,verma2014synchronization,verma2013potential,verma2015kuramoto,bera2017coexisting,boccaletti2018synchronization}, extreme events~\cite{ansmann2013extreme, chowdhury2019synchronization}, and chimera states~\cite{parastesh2021chimeras,sharma2021chimeralike,kumar2017partially}, that individual units of the collective are devoid of. In recent years, several theoretical and experimental investigations have established \textit{oscillation quenching}~\cite{sharma2012amplitude,verma2017explosive,saxena2012amplitude,koseska2013oscillation,zou2021quenching,toiya2008diffusively,suresh2014experimental} 
 and \textit{oscillation revival}~\cite{zou2019quenching,yadav2018revival, zou2017revival, senthilkumar2016experimental,miake2002biological} i.e. the restoration of activity in quenched oscillators,  as robust emergent phenomena in varied complex systems. On the other hand, rhythmogenesis~\cite{mandal2013conjugate,parmananda2001resonance,chigwada2006resonance,kumar2016resonance,montoya2014reversing} - mechanisms to stimulate oscillations from stable fixed point states have been also garnering interest. These collective phenomena provide mechanisms to control the dynamics of the constituent subsystems through specific coupling forms.
 
 Control mechanisms enable stabilization of dynamical systems to desired asymptotic states~\cite{pisarchik2014control}. Precise control of nonlinear systems~\cite{sevilla2017error,sevilla2015selective}, especially chaotic systems has been an area of intense research~\cite{boccaletti2000control} due to its vast potential for engineering applications. Control methods, such as the OGY method~\cite{ott1990erratum} and threshold control~\cite{sinha1998dynamics,sinha1993adaptive,murali2003experimental} have been developed to curtail the activity in the chaotic systems to obtain desirable states. On the contrary, methods to induce activity and preserve complex dynamics in systems, termed as the anticontrol of chaos, have also been widely worked on~\cite{yang1995preserving,chen1997control,chen2006introduction}. 

It is important to note that the aforementioned schemes to regulate dynamics, are typically applicable in lone dynamical systems. Therefore, it is imperative to design a method to yield any target dynamics in the constituent units of a coupled system. A plausible approach is to control the shared information between the interacting nodes~\cite{chowdhury2019convergence,ghosh2022synchronized} in the schemes that attain oscillation quenching or rhythmogenesis. In this letter, we introduce an \textit{intermittent control} approach to acquire the desired attractor, where an interaction function changes dynamically as a function of time or the states of the individual oscillators. The intermittent interaction used in this work maybe comparable to other schemes like finite time step method~\cite{amritkar1993synchronization}, dynamic coupling~\cite{schrodertransient,dixit2021emergent,aprasad_pramana, yadav2017dynamics, sudhashu2019, threshold_2019,dixit2020static,asir2021critical,dixit2021dynamic,shajan2021enhanced}, sporadic coupling~\cite{zochowski2000intermittent}, periodic coupling~\cite{li2018network} and stochastic on-off coupling~\cite{jeter2015synchronization}. The term `intermittent' is used here to indicate the discontinuous nature of the interaction between the coupled systems. The prime focus of this work is to recast intermittent coupling that can yield both oscillation suppression and rhythmogenesis, as a powerful leash to control the dynamics of coupled subsystems to desired stable states.

Systems where interactions get activated as a function of time or the system state abound in both natural and engineered contexts.  In biological systems, gene regulation is dynamically controlled as a function of system state~\cite{nie2017estimating}. In robotic networks with swarms of robots, interactions occur only within a limited range in physical space~\cite{kantaros2015distributed}. In ocean exploration~\cite{leonard2007collective}, the robots typically communicate on aquatic acoustic channels over long distances that are susceptible to environmental interference. In these applications, maintaining constant connections are energy expensive and efficient intermittent coupling schemes that maximise control and minimize information transfer are highly sought after. It is also worth noting that this intermittent scheme opens up the possibility to control the dynamics of the systems whose parameters and coupling strengths are fixed and cannot be modified continuously.

In coupled systems, both oscillation suppression and rhythmogenesis have been widely explored as important dynamical phenomena. Yet a simple control scheme that utilizes this suppression and stimulation of activity to access any desired dynamical state (from steady states to chaos) has not been proposed. We achieve this fine control by limiting the interaction among the subsystems through dynamic modulation (switching ON and OFF) of the coupling term. This is experimentally demonstrated in a pair of Chua's circuits both as a function of time and state variables. In the next section, this intermittent control scheme is described in detail. 

\section{Scheme}\label{scheme}
	
Consider $N$($=2$) identical $m$-dimensional nonlinear oscillators coupled via dynamic interaction in the following two ways: 

	\begin{equation}
\dot{\mathbf{X}}_{i} ={F}(\mathbf{X}_{i})+ \epsilon \gamma \sum_{j=1}^{N}  ({H} \mathbf{X_j} - \mathbf{X}_{i}),
	\label{eq1}
	\end{equation}
and 
	\begin{equation}
\dot{\mathbf{X}}_{i} ={F}(\mathbf{X}_{i})- \epsilon \gamma H \sum_{j=1}^{N}  (\mathbf{X_j} - \mathbf{X}_{i}),
	\label{eq2}
	\end{equation}

where, $i,j=1,2$, and $i \neq j$. $\mathbf{X}_{i}$ represents state variables of the $m$-dimensional $i$-th oscillator and $F: \mathbb{R}^m \to \mathbb{R}^m$ is the vector field describing its intrinsic dynamics. $\gamma$ is an $m\times m$ diagonal matrix that determines which components of $X_i$ take part in the dynamic coupling. The interaction strength $\epsilon\geq 0$, is taken to be identical for all oscillators. Equation \ref{eq1} represents an \textit{attractive interaction}, where the state variables $\mathbf{X}_{1}$ and $\mathbf{X}_{2}$ tend to approach each other under the influence of coupling. Equation \ref{eq2} represents a \textit{repulsive interaction} where the state variables $\mathbf{X}_{1}$ and $\mathbf{X}_{2}$ repel each other~\cite{majhi2020perspective,aravind2021competitive,dixit2019dynamics,dixit2020static,hong2011kuramoto,mishra2015chimeralike,aravind2021emergent}. 

Dynamic control of the coupling terms is achieved through the step function $H$~\cite{aprasad_pramana, schrodertransient,dixit2021dynamic,dixit2021emergent}. $H(\mathbf{X}_{i},t)$ takes the values $0$ or $1$ either as a function of time $t$ or the state variables $\mathbf{X}_{i}$. When $H=1$ the two oscillators are coupled to each other and when $H=0$ they are completely isolated from each others' influence. When $H$ depends on the state variables of the two oscillators, i.e.,

\begin{equation}
H(\mathbf{X}_{i},t)=\begin{cases}
          1 \quad &\text{if} \, \mathbf{X}_{i} \in R' \\
          0 \quad &\text{if} \, \mathbf{X}_{i} \notin R' \\
     \end{cases}
\label{Hx}
\end{equation}

this yields a \textit{state dependant interaction}. Where $R'$ is a subset of the state space $\mathbb{R}^m$ where the interaction is active. A measure of this subset is obtained from the normalized fraction $\Delta = \Delta'/\Delta_a $, where $\Delta_a$ is the width of the uncoupled attractor along the direction of the coupled state variable and $\Delta'$ is the width of the region in which the coupling is active. \textit{Time dependant interaction} is when $H$ is explicitly dependant on time. For instance $H$ can be a periodic step function of time period $T$ as follows, 

\begin{equation}
H(\mathbf{X}_{i},t)=\begin{cases}
          1 \quad &\text{if} \, 0< t \leq \tau'  \\
          0 \quad &\text{if} \, \tau'< t \leq T \\
     \end{cases}
     \label{Ht}
\end{equation}

\noindent Here, $\tau=\tau'/T$ is a measure for the fraction of time the interaction is active.

These two parameters $\Delta$ and $\tau$ that control the degree of interaction between the coupled systems, allow us to harness a given coupled oscillator into desired stable states. Specifically, the attractive interaction (c.f. Eq.~\ref{eq1}) induces suppression of activity among the subsystems and the repulsive interaction (c.f. Eq.~\ref{eq2}) promotes the revival of activity~\cite{chen2009dynamics, liang2012phase}. This scheme therefore enables control and anticontrol of chaos in the constituent oscillators of a coupled system.  Given this framework, we experimentally demonstrate this scheme in a pair of Chua's circuits. 

\begin{figure*}
\includegraphics[width=0.95\textwidth]{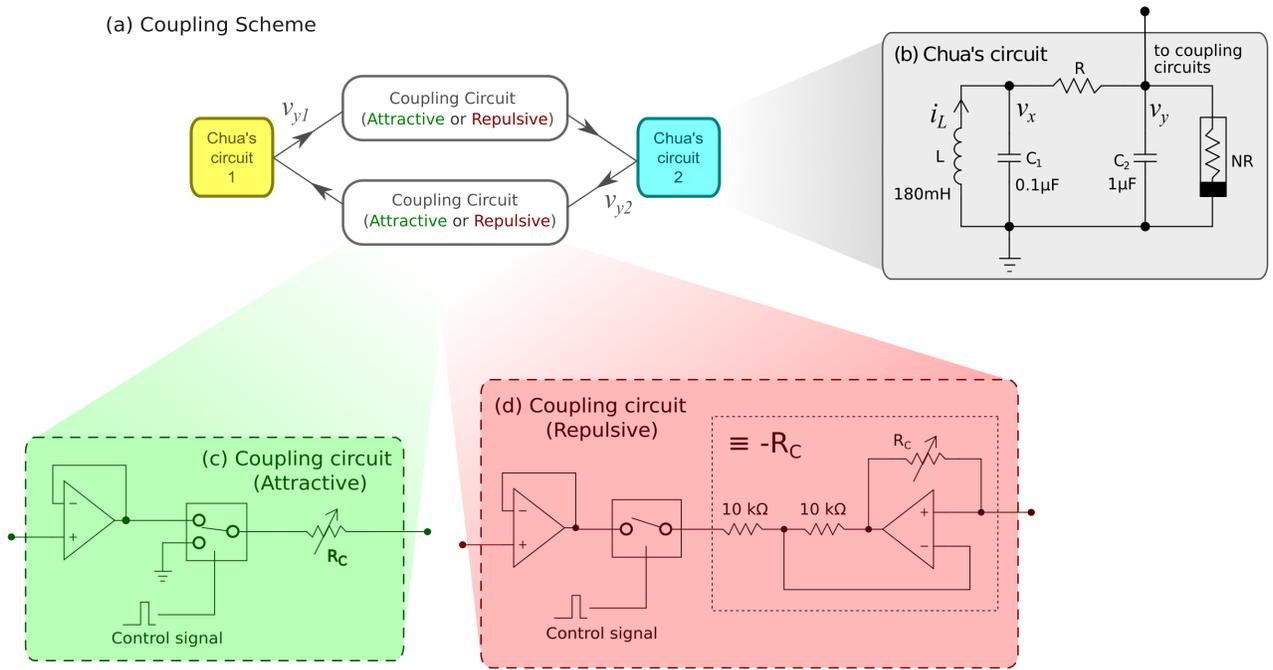}
\caption{A schematic representation  of the electronic circuit implementation of the control scheme. (a) Block diagram depicting two Chua's circuits coupled through the y-variables. Bidirectional coupling is established using two coupling circuits (one in each direction). The attractive coupling circuit yields the coupling form in Eq.~\ref{eq1} and the repulsive coupling circuit yields the form in Eq.~\ref{eq2}. (b) Chua's circuit with values of all components used. The parameter values of the nonlinear resistor ($N_R$) are $G_a = -0.756~m\Omega^{-1}$, $G_b = -0.401~m\Omega^{-1}$ and $B_P = \pm 1.08~V$. The voltages $v_x$, $v_y$ (across capacitors $C_1$ and $C_2$) and the current $i_L$ (through the inductor $L$) are proportional to the state variables $x$, $y$ and $z$ respectively. (c) The coupling circuit used to implement the attractive coupling form from Eq.~\ref{eq1}. Two of these coupling circuits are used to establish bidirectional coupling between the two Chua's circuits. (d) Coupling circuit used to implement the repulsive coupling form described by Eq.~\ref{eq2}. A negative resistance of $-R_c$ was obtained using a simple negative impedance converter shown within the dotted lines in the figure.}
\label{circuit}
\end{figure*}

\begin{figure*}
\includegraphics[width=\textwidth]{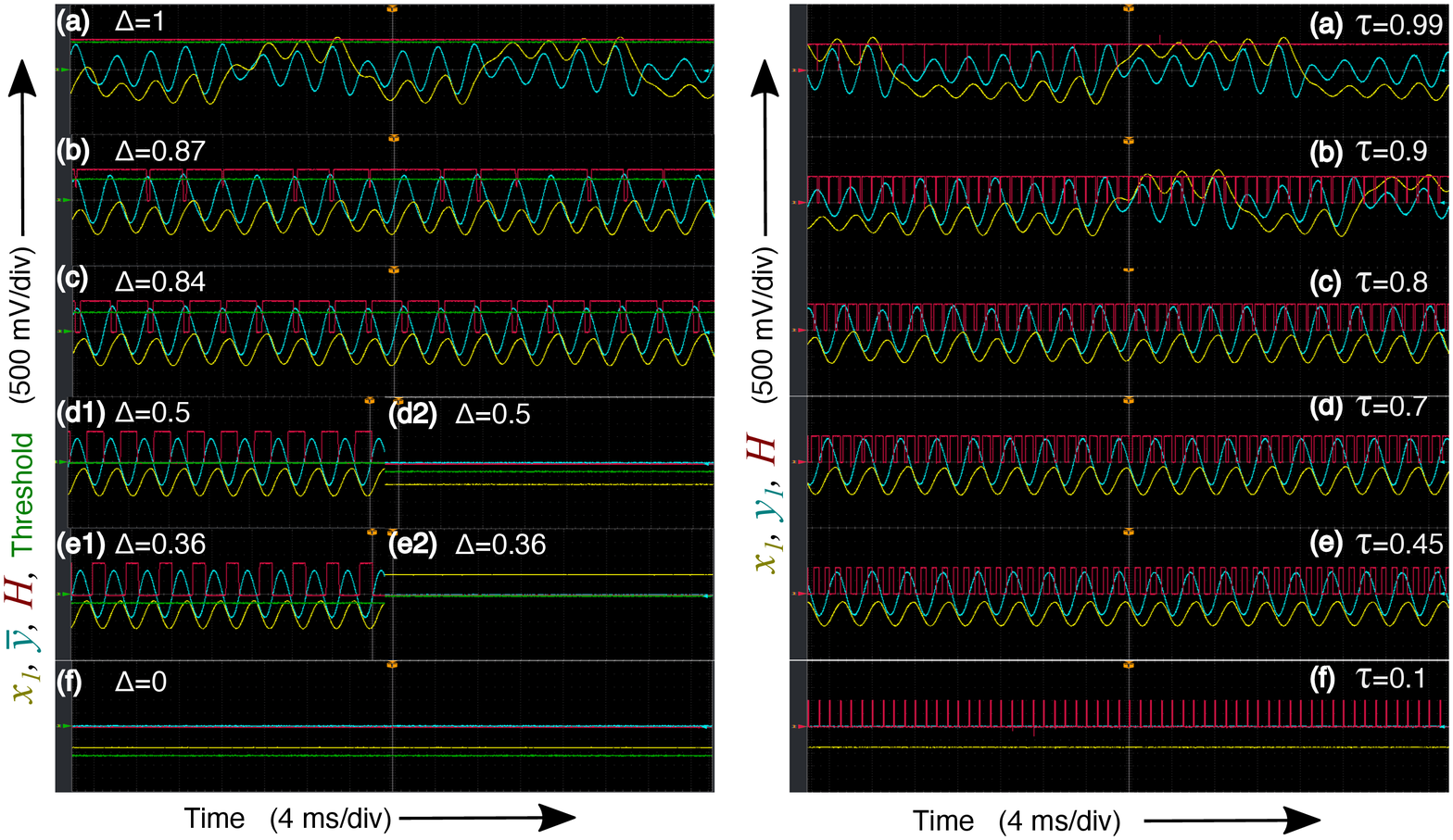}
\caption{(Color online.) Left: \textit{State dependant interaction} (a) Oscilloscope traces of $v_{x1}$ (yellow), $\bar{v_y}$(blue), $H$(red) and (threshold voltage)(green). The coupling resistance is fixed at $R_C = 4380~\Omega$ and resistance $R$ is fixed at $R = 1750~\Omega$. The specific parameter values corresponding to each trail from top to bottom are as follows: (a) $\Delta=1$ (b) $\Delta=0.87$ (c) $\Delta=0.84$ (d1) $\Delta=0.5$ (d2) $\Delta=0.5$ (e1) $\Delta=0.36$ (e2) $\Delta=0.36$ (f) $\Delta=0$. Here $\Delta$ denotes the fraction of state space where the interaction is active (c.f. Eq~\ref{Hx}). As $\Delta$ is reduced we see a systematic reduction in the activity of the subsystems. Right: \textit{Time dependant interaction} Oscilloscope traces of $v_{x1}$(yellow), $v_{y1}$(blue) and $H$(red). The specific parameter values corresponding to each trail from top to bottom are as follows: (a) $\tau=0.99$ (b) $\tau=0.9$ (c) $\tau=0.8$ (d) $\tau=0.6$ (e) $\tau=0.3$ (f) $\tau=0.1$. Here $\tau$ (dutycycle of the periodic control signal) denotes the interaction active time described in Eq.~\ref{Ht}. Again, for decreasing values of $\tau$ a systematic transition from chaotic, to periodic, to fixed point is observed.}
\label{tseries}
\end{figure*}

\section{Experimental Implementation}
\label{imple}

The control schemes described above were implemented in a pair of coupled Chua's circuits. A schematic representation of the experimental setup is detailed in Fig.~\ref{circuit}. All components used in the construction are mentioned explicitly in the same. We first consider the attractive interaction described by Equation~\ref{eq1}. The governing equations of the coupled circuit (c.f. Fig.~\ref{circuit}) attain the form in Eq.~\ref{eq1} where,

\begin{equation}
F = 
\begin{bmatrix}
\alpha(y-x-g(x))\\
x-y+z\\
-\beta y
\end{bmatrix}
\
\gamma = 
\begin{bmatrix}
0 & 0 & 0\\
0 & 1 & 0\\
0 & 0 & 0
\end{bmatrix}
\label{eq5}
\end{equation}

\begin{equation}
    g(x) = bx + \frac{1}{2} (b-a)(|x-1|-|x+1|)
    \label{eq6}
\end{equation}

\noindent The circuit equations are nondimensionalized as follows, 

\begin{eqnarray*}
x = \frac{v_x}{B_P},  \ \ y = \frac{v_y}{B_P}, \ \ z= \frac{i_L R}{B_P}, \ \ t = \frac{t'}{C_2 R},
\\
\alpha = \frac{C_2}{C_1}, \ \ \beta = \frac{C_2R^2}{L}, \ \ \epsilon = \frac{R}{R_C},
\\
a = G_a R, \ \ b = G_b R.
\end{eqnarray*}

The coupling between the two Chua subsystems is via the $y$ variable (proportional to voltage $v_y$ across capacitor $C_2$). For the case of \textit{state dependant interaction} (c.f. Eq.~\ref{Hx}), a threshold is set on the mean voltage of the coupled variable, i.e., when $\overline{v_y}$ is above a threshold value, coupling is OFF ($H=0$). Changing this threshold voltage allows fraction $\Delta$ to be manipulated. Such a state dependant intermittent coupling is experimentally implemented with two one-way coupling circuits depicted in Fig.~\ref{circuit}c. A voltage controlled single pole double throw (SPDT) switch is implemented using analog CMOS switches (AD7510DI). This switch is controlled using the output of a comparator that monitors the mean value of $v_y$ and compares it with a fixed \textit{threshold voltage}. Thus the coupling dynamically switches ON and OFF ($H = 1$ or $0$) based on the live values of the system variables $v_{y1}$ and $v_{y2}$ of the subsystems. 

Oscilloscope trails obtained from this experimental implementation is shown on the left part of Fig.~\ref{tseries}, voltage $v_{x1}$ of Chua $1$(yellow), mean $\overline{v_y}$ (blue), interaction control $H$ (red) and threshold voltage level (green) are presented. The resistance $R$ is fixed at $R = 1750~\Omega$ where the uncoupled systems exhibit double-scroll chaos. The control $H$ is ON only when $\overline{v_y}$ is below the threshold. In the top panel the mean $\overline{v_y}$ (blue) is always below the threshold level (green). Hence the control $H$ is always ON and $\Delta = 1$. As we decrease the threshold, control term $H$ goes to $0$ (OFF) every time $\overline{v_y}$ crosses the threshold. This leaves just a negative feedback ($-\epsilon \mathbf{X}_{i}$) term in equation \ref{eq1} thereby suppressing oscillations. As the threshold reduces, $\Delta$ becomes smaller and interaction active region reduces in size. This simultaneously leads to oscillation suppression. From panels (a) to (f) we see a systematic control of the dynamics reminiscent of inverse period doubling as the oscillations completely die down for small values of $\Delta$. Panels (d) and (e) also show the coexistence of both steady state and limit cycle behaviour for the same values of $\Delta$. The existence of this bistable regime evokes the possibility of hysteresis with respect to $\Delta$. This is numerically explored in the Supplemental Material \cite{supplementary}.

\textit{Time dependant interaction} (c.f. Eq.~\ref{Ht}) that yields controlled suppression of oscillations in the subsystems can be implemented using the same coupling circuit given in Fig.~\ref{circuit}(c), where the control signal is an externally generated square wave. The duty cycle of this control signal is analogous to the fraction $\tau$ described in Eq.~\ref{Ht}. This fixes the fraction of time for which the interaction is ON ($H=1$). The control signal used in this implementation is a square wave oscillating between 0~V and 2~V at a frequency of 1 KHz. 

On the right part of Fig.~\ref{tseries} oscilloscope trails from the implementation of \textit{time dependant interaction} are depicted. The control signal $H$ is shown in red along with the state variables $v_{x1}$ (yellow) and $v_{y1}$ (blue) of Chua 1 (c.f. Fig.~\ref{circuit}). Here we see $H$ is a periodic signal and interaction time $\tau$ (duty cycle) systematically reduced from top to bottom. This is accompanied by a systematic suppression of activity as the system transitions from chaotic to periodic to oscillation death states.

The repulsive coupling form described by equation~\ref{eq2} was electronically implemented as shown in Fig.~\ref{circuit}(d). The same voltage controlled CMOS switch (AD7510DI) was used in the SPST configuration intermittently couple the two subsystems. The negative resistance necessary to to implement a $-\epsilon$ was implemented using a negative impedance converter, shown within the dotted box in Fig.~\ref{circuit}(d). The oscilloscope timetrails depicting stimulation of oscillations from fixed points have been shown in the Supplemental Material \cite{supplementary}.

The scheme detailed in section II was thus experimentally demonstrated using coupled Chua's circuits. Now, the behaviour of this coupled system is studied in detail using both experiments and numerical simulations over a range of coupling parameters. This establishes the viability of dynamically controlled interactions as a control mechanism for coupled dynamical systems. The four different coupling regimes that include attractive and repulsive interactions controlled by both state and time are examined case by case. 

\section{Numerical Exploration}
\label{results}

\subsection{Suppression of oscillations}
\begin{figure}
\includegraphics[width=0.48\textwidth]{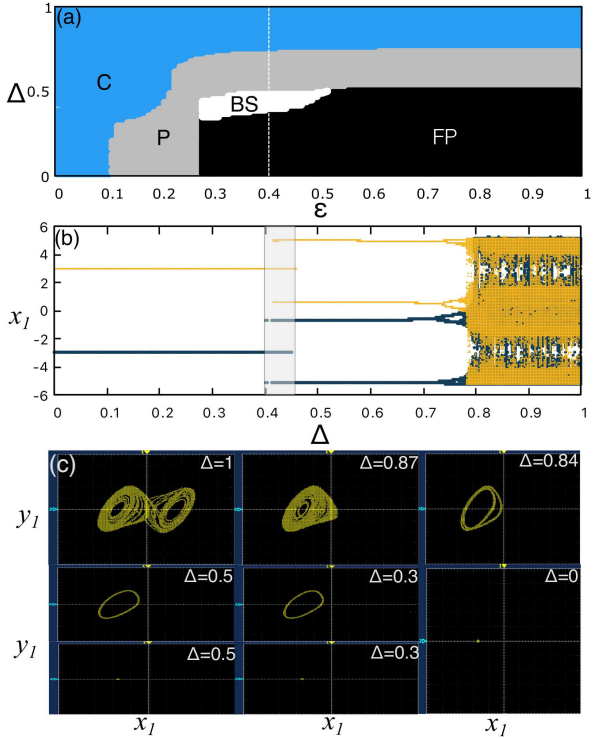}
\caption{ Suppression through space-dependant interaction: (a) Different dynamical states of two coupled Chua oscillators (c.f. Eqs.~\ref{eq1} \& \ref{eq5}) in the parameter plane $(\epsilon-\Delta)$. The regimes marked C, P, FP, and BS represent chaotic, periodic, fixed point and bistable (co-existence of oscillatory and fixed point state) state respectively. (b) Bifurcation diagram of the coupled Chua system (c.f. Eqs.~\ref{eq1} \& \ref{eq5}) is plotted with interaction active state space $\Delta$ at $\epsilon=0.41$, obtained by sampling the relative maxima and minima of the time history of $x_1(t)$. The different colors show attractors corresponding to positive and negative initial conditions. (c) Experimental phase portraits for $\epsilon=0.41$ corresponding to various values of the control parameter $\Delta$. The time trails corresponding to the same attractors were shown in Fig.~\ref{tseries}}
\label{fig3}
\end{figure}
\begin{figure}
\includegraphics[width=0.48\textwidth]{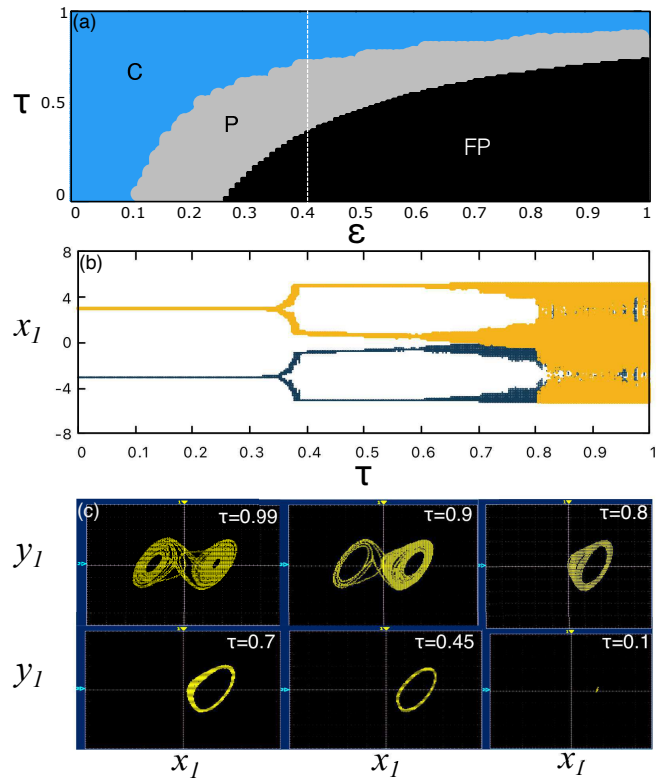}
\caption{Suppression through time-dependant interaction: (a) Different dynamical states of two coupled Chua oscillators (c.f. Eqs.~\ref{eq1} \& \ref{eq5}) in the parameter plane $(\epsilon-\tau)$. The regimes marked C, P and FP represent chaotic, periodic and fixed point state respectively. (b) Bifurcation diagram of the coupled Chua system (c.f. Eqs.~\ref{eq1} \& \ref{eq5}) is plotted with interaction active time $\tau$ at $\epsilon=0.41$, obtained by sampling the relative maxima and minima of the time history of $x_1(t)$. The different colors show attractors corresponding to positive and negative initial conditions. Experimental phase portraits for $\epsilon=0.41$ corresponding to various values of the control parameter $\tau$. The time trails corresponding to the same attractors were shown in Fig.~\ref{tseries}}
\label{fig4}
\end{figure}

First, the attractive coupling scheme described by Eq.~\ref{eq1} is studied. We explore the behaviour of this coupled system when the interactions are controlled by the \textit{current state} of the coupled system. The control $H$ becomes zero when the mean $\bar{y}$ becomes lesser than some threshold. Numerical simulation of this coupled Chua's system (refer Eq.~\ref{eq5}) were carried out for $R=1850~\Omega$ (double scroll chaos) for a range of coupling strengths $\epsilon$ and control parameter $\Delta$. In Figure~\ref{fig3} the top panel depicts and classifies the behaviour of the coupled Chua's systems in the space of aforementioned parameters ($\Delta$ and $\epsilon$). We see large regions of chaotic (C), periodic (P) and fixed point (FP) regimes. The small white region in the centre shows coupling parameters for which steady states and oscillatory states coexist. The core strength of this controlled scheme is the fact that all these states can be accessed simply by changing the amount of intermittent interaction (through $\Delta$) without even necessitating a change in coupling strength $\epsilon$. The middle panel shows clear bifurcations in system dynamics for a fixed $\epsilon = 0.4$ as a function of $\Delta$. As $\Delta$ decreases, the activity of the subsystems faithfully suppress from double scroll chaos to single scroll chaos to limit cycles and fixed points. The last panel shows oscilloscope snapshots depicting phase space portraits from the experimental implementation corresponding to same control regime for $\epsilon = 0.4$. Note that the attractors correspond to the time trails shown on the left of Fig.~\ref{tseries}. We clearly see, decreasing $\Delta$ offers a robust method to regulate the dynamics of coupled systems. 

Figure~\ref{fig4} corresponds to the attractive coupling scheme controlled by the an external time dependant control signal. Here, as detailed earlier, a periodic step function of time period T comparable to the natural frequency of the uncoupled oscillator has been used. The fraction of time that the interaction is ON ($\tau$) acts as a robust handle to control the activity of the systems to the desired steady state. Top panel shows the plethora of stable behaviors accessible through this mechanism. the middle panel depicts bifurcation of system dynamics from fixed points to double scroll chaos as a function of the interaction time $\tau$. The last panel portrays, experimental attractors obtained from the coupled Chua's circuits for a range of $\tau$ values. These correspond to the time trails shown in the right of Fig.~\ref{tseries}. We see granular control in behaviour as a  function of $\tau$. For a given coupling strength $\epsilon$, note we can change from one dynamical state to another by varying the fraction $\tau$.

\subsection{Stimulation of activity}

\begin{figure}
\includegraphics[width=0.48\textwidth]{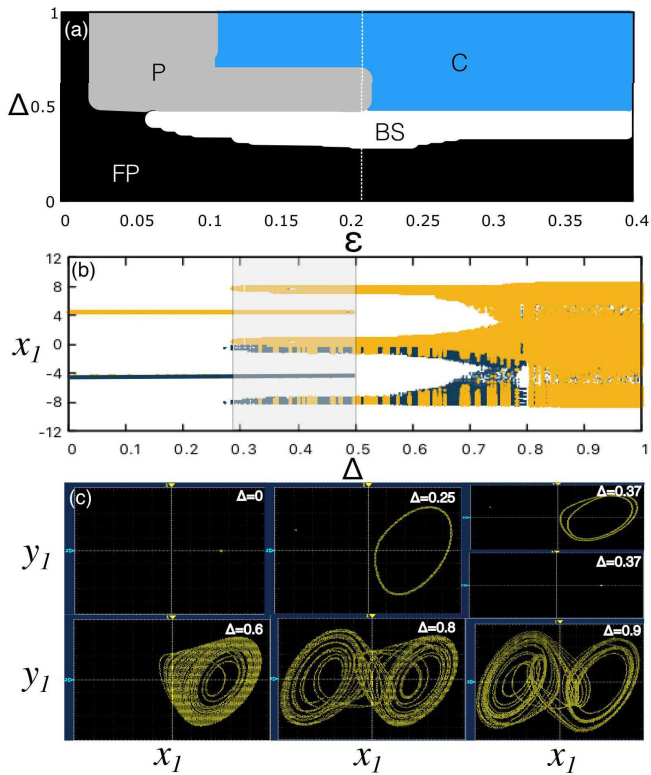}
\caption{Stimulation through space-dependant interaction: (a) Different dynamical states of two coupled Chua oscillators (c.f. Eqs.~\ref{eq2} \& \ref{eq5}) in the parameter plane $(\epsilon-\Delta)$. The regimes marked C, P, FP and BS represent chaotic, periodic, fixed point and bistable (co-existence of oscillatory and fixed point state) state, respectively. (b) Bifurcation diagram of the coupled Chua system (c.f. Eqs.~\ref{eq2} \& \ref{eq5}) is plotted with interaction active state space $\Delta$ at $\epsilon=0.21$, obtained by sampling the relative maxima and minima of the time history of $x_1(t)$. The different colors show attractors corresponding to positive and negative initial conditions. Experimental phase portraits for $\epsilon=0.21$ corresponding to various values of the control parameter $\tau$. The time trails corresponding to the these attractors can be found in the Supplemental Material \cite{supplementary}.}
\label{fig5}
\end{figure}

\begin{figure}
\includegraphics[width=0.48\textwidth]{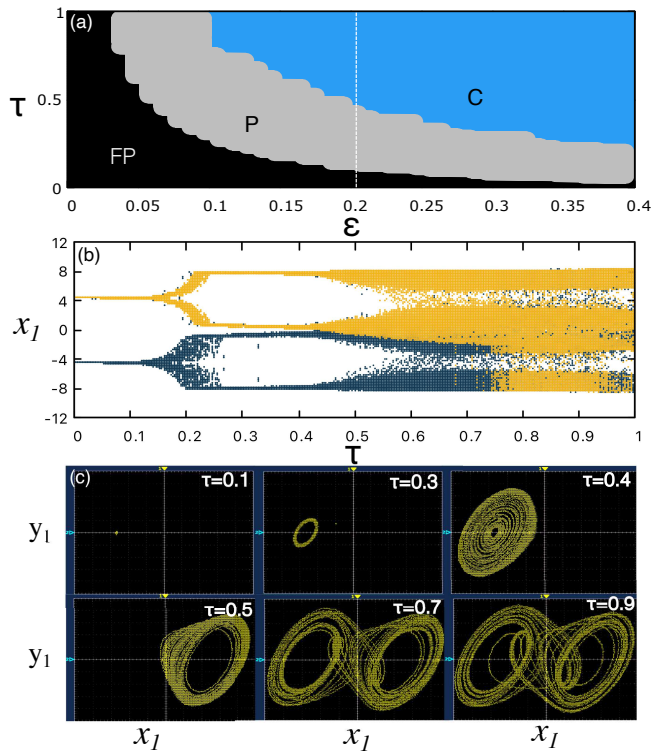}
\caption{Stimulation through time-dependant interaction: (a) Different dynamical states of two coupled Chua oscillators (c.f. Eqs.~\ref{eq2} \& \ref{eq5}) in the parameter plane $(\epsilon-\tau)$. The regimes marked C, P and FP represent chaotic, periodic and fixed point state, respectively. (b) Bifurcation diagram of the coupled Chua system (c.f. Eqs.~\ref{eq2} \& \ref{eq5}) is plotted with interaction active time $\tau$ at $\epsilon=0.21$, obtained by sampling the relative maxima and minima of the time history of $x_1(t)$. The different colors show attractors corresponding to positive and negative initial conditions.  Experimental phase portraits for $\epsilon=0.21$ corresponding to various values of the control parameter $\tau$. The time trails corresponding to the these attractors can be found in the Supplemental Material \cite{supplementary}.}
\label{fig6}
\end{figure}

Next, the repulsive coupling scheme described in Eq.~\ref{eq2} was implemented. The time dependant control mechanisms remain the same. Repulsive coupling, specifically the form considered in Eq.~\ref{eq2} is known to revive oscillations~\cite{chen2009dynamics}. By introducing intermittency in the coupling term precise control over the degree of revival is possible. This is demonstrated both in simulation and experiments. The experimental implementation of a coupling with a negative coupling strength ($-\epsilon$) was made possible through a simple negative impedance converter (refer Fig~\ref{circuit}(d)). Bidirectional repulsive coupling is therefore achieved by using two unidirectional coupling circuits as depicted in Fig.~\ref{circuit}. Oscilloscope trails obtained from the experimental implementation is shown in the supplementary material.

Figure~\ref{fig5} details the case of \textit{state dependant interaction} when the nature of the coupling is repulsive. The parameters of the subsystems are fixed such that the uncoupled systems exhibit fixed point dynamics. So for the simulations in Fig.~\ref{fig5} $R = 2050~\Omega$ and for the experimental implementation $R$ was fixed at $1950~\Omega$. The top panel shows regions in $\Delta$ and $\epsilon$ where the subsystems exhibit chaotic (C), fixed point (FP), periodic (P) and bistable (BS) dynamics. In the middle panel, a bifurcation along increasing $\Delta$ at $\epsilon = 0.21$ reveals a systematic progression to double scroll chaos. Note that the attractors obtained by exciting fixed points are much larger than those in Chua's circuit whose native parameters are in the chaotic regime (compare with Figs.~\ref{fig3} and \ref{fig4}). A large bistable region where both fixed point and limit cycle solutions coexist was also found. This occurs as space dependant coupling introduces strong initial condition dependence in these regions. The last panel from experiments, clearly portrays the controlled revival of oscillation to any desired degree and the bistable regions for middle values of $\Delta$ in close correlation with numerical findings. Finally, Figure~\ref{fig6} considers the case of \textit{time dependant interaction} i.e. repulsive interactions controlled by time varying signals. Top panel shows three separated regions of behaviour with no multistability. The middle panel depicts a clear increase in complexity with increasing $\tau$. The experimental phase portraits closely match the numerically obtained bifurcation results.  

\section{Conclusion}\label{conclusion}

A robust scheme to control the dynamics of constituent oscillators in coupled systems was proposed. The scheme was experimentally demonstrated using coupled Chua's circuits with dynamic interactions facilitated through controllable coupling circuits. 
We show that limiting interaction between coupled oscillators in a controlled fashion by introducing intermittency allows for a precise control of the constituent oscillators, without access to either the system parameters, nor the strength of the interaction. This may lead to novel control strategies of coupled networks and may allow the harnessing of these predictable dynamics for engineering applications. The effectiveness of this control mechanism was demonstrated in a prototypical coupled chaotic system (Chua's circuits) both numerically and experimentally. This suggests that stroboscopic modulation of the interaction term as a function of state or time, yields a simple and potent mechanism for accessing the entire spectrum of behaviours in coupled chaotic systems.

Two new parameters $\Delta$ and $\tau$ have been introduced that limit the amount of interaction among the coupled subsystems by controlling the fraction of state space or time period that the interaction is ON. These parameters provide handles to realize granular control of dynamics of the subsystems, while no direct access to the system state, parameters or coupling strength is available. Thus this experimental implementation may serve as a general template based on which system specific adaptations of this approach maybe realized. 

Two coupling forms (attractive and repulsive) were demonstrated in this work. The attractive form suppresses the activity in subsystems of the coupled network and the repulsive form stimulates activity in the subsystems. These two forms can thereby be considered as coupled system analogues to methods that yield control and anticontrol of chaos. 

This intermittent control scheme robustly decides the asymptotic states of self-excited systems (such as Chua's circuit). Thus, it will also be interesting to see a variant of this intermittent control scheme applied to systems with hidden attractors whose basin do not intersect with any unstable fixed point~\cite{dudkowski2016hidden}.We would also like to emphasize that this proposed intermittent control technique may be applied in various other experimental contexts, where precise control of constituent oscillator dynamics is required of coupled systems like mercury beating heart (MBH) oscillators, electrochemical systems, mechanical oscillators and optoelectronic systems.

This work illustrates this idea of control through intermittent interactions in a minimal setting of two oscillators. However the general principles of oscillator suppression and revival described here are number independent. A natural progression of this work would be to realize such limited interaction based control in larger number of coupled units. Given the occurrence of bistable regimes in state dependant interaction schemes, a large network coupled intermittently may give rise to chimera states that can be effectively controlled using the same principles. 
Hence, this control scheme may serve as a powerful tool regulate dynamics in a wide variety of contexts.

\end{document}